# An Optimal Load-Frequency Control Method for Inverter-Based Renewable Energy Transmission

Kirsch Mackey and Roy A. McCann
Department of Electrical Engineering
University of Arkansas
Fayetteville, AR USA
knmackey@uark.edu

***Abstract*—The frequency droop response of conventional turbine driven synchronous generators with respect to load increases is normally used in order to have stable operating characteristics for multiple generators operating in parallel over large geographical regions. This presents a challenge for renewable energy sources that interface to the transmission grid through static inverters that do not exhibit an intrinsic frequency droop characteristic. This paper provides a technique for designing optimal load frequency controllers as transmission line inverters fed from renewable energy sources that allows for fast dynamic response due to variable solar and wind conditions while maintain stability to interconnected synchronous generators. A control technique based on LQG optimization theory is presented. Detailed analysis of a three-area system in a region of mixed wind and solar photovoltaic sources is modeled in a manner that confirms the effectiveness of the disclosed load-frequency control method.**

## I. Introduction

Increasing use of renewable energy sources such as wind and solar often requires transmission of electricity over distances where complex inter-area power flow interactions occur with conventional generation. This is due in part to the abundance of renewable energy sources in rural areas, such as the case of the central and southwestern United States that are far removed from major population centers. Whenever EHV transmission line lengths exceed 250 km then inter-area frequency stability becomes a concern. In the case of conventional turbine0driven synchronous generators, th9ere is an extensive theory for developing feedback controllers to ensure stable operation of the transmission grid as described in [1] and [2]. For conventional synchronous generators, a frequency-droop characteristic is designed such that frequency decreases as load increases, thereby leading to load sharing between parallel connected machines. However, for static inverters commonly used for interfacing renewable electric power sources such as photovoltaics and wind, the power electronics has no such intrinsic response. This can lead to stability in power flow control and load sharing between renewable and conventional generation as described in [3], [4] and [5]. This has motivated the development of solutions such as those found in [6] and [7]. However, a virtual-droop characteristic can be programmed as part of the inverter operation that approximates the load sharing properties of synchronous generators. This paper considers a stochastic approach that incorporates the probabilistic nature of wind and solar sources to design the controller coefficients. In addition, the availability of synchrophasors along with smart grid technologies enables the communication of angle and frequency measurements with minimal time-delay. That is, the research proposes that stochastic optimal control theory can be applied to improve the stability performance of interconnected renewable energy sources in order to achieve proper load-frequency regulation objectives.

## II. Modeling of Inter-Area Dynamics

The following modeling of inter-area dynamics used in this research is developed in the manner detailed by Murty in [8]. In this research the time-constants and parameters are modified to match the faster response times associated with static inverters as described by [9]. This paper considers a three-area load-frequency control system. The most general case allows for inter-tie power flows between any two areas. This is shown in terms of Laplace-domain block diagram form in Fig. 1. The corresponding transfer functions are given in [8] by,

$$\Delta F_1(s) = \frac{K_{P1}}{1 + sT_{P_1}}\left[\Delta P_{G1} - \Delta P_{D1} - \Delta P_{tie,1}\right] \tag{1}$$

$$\Delta F_2(s) = \frac{K_{P2}}{1 + sT_{P_2}}\left[\Delta P_{G2} - \Delta P_{D2} - \Delta P_{tie,2}\right] \tag{2}$$

$$\Delta F_3(s) = \frac{K_{P3}}{1 + sT_{P_3}}\left[\Delta P_{G3} - \Delta P_{D3} - \Delta P_{tie,3}\right] \tag{3}$$

$$\Delta S_{V1}(s) = \frac{1}{1 + sT_{S_1}}\left[\Delta P_{C1}(s) - \frac{1}{R_1}\Delta F_1(s)\right] \tag{4}$$

$$\Delta S_{V2}(s) = \frac{1}{1 + sT_{S_2}}\left[\Delta P_{C2}(s) - \frac{1}{R_2}\Delta F_2(s)\right] \tag{5}$$

$$\Delta S_{V3}(s) = \frac{1}{1 + sT_{S_3}}\left[\Delta P_{C3}(s) - \frac{1}{R_3}\Delta F_3(s)\right] \tag{6}$$

$$\Delta P_{G1}(s) = \frac{1}{1 + sT_{TG1}}\Delta X_{V1}(s) \tag{7}$$

$$\Delta P_{G2}(s) = \frac{1}{1 + sT_{TG2}}\Delta X_{V2}(s) \tag{8}$$

$$\Delta P_{G3}(s) = \frac{1}{1 + sT_{TG3}}\Delta X_{V3}(s) \tag{9}$$

$$\Delta P_{tie,1} = \frac{2\pi T_{12}^0}{s}\left[\Delta F_1(s) - \Delta F_2(s)\right] + \frac{2\pi T_{13}^0}{s}\left[\Delta F_1(s) - \Delta F_3(s)\right] \tag{10}$$

$$\Delta P_{tie,2} = \frac{2\pi T_{21}^0}{s}\left[\Delta F_2(s) - \Delta F_1(s)\right] + \frac{2\pi T_{23}^0}{s}\left[\Delta F_2(s) - \Delta F_3(s)\right] \tag{11}$$

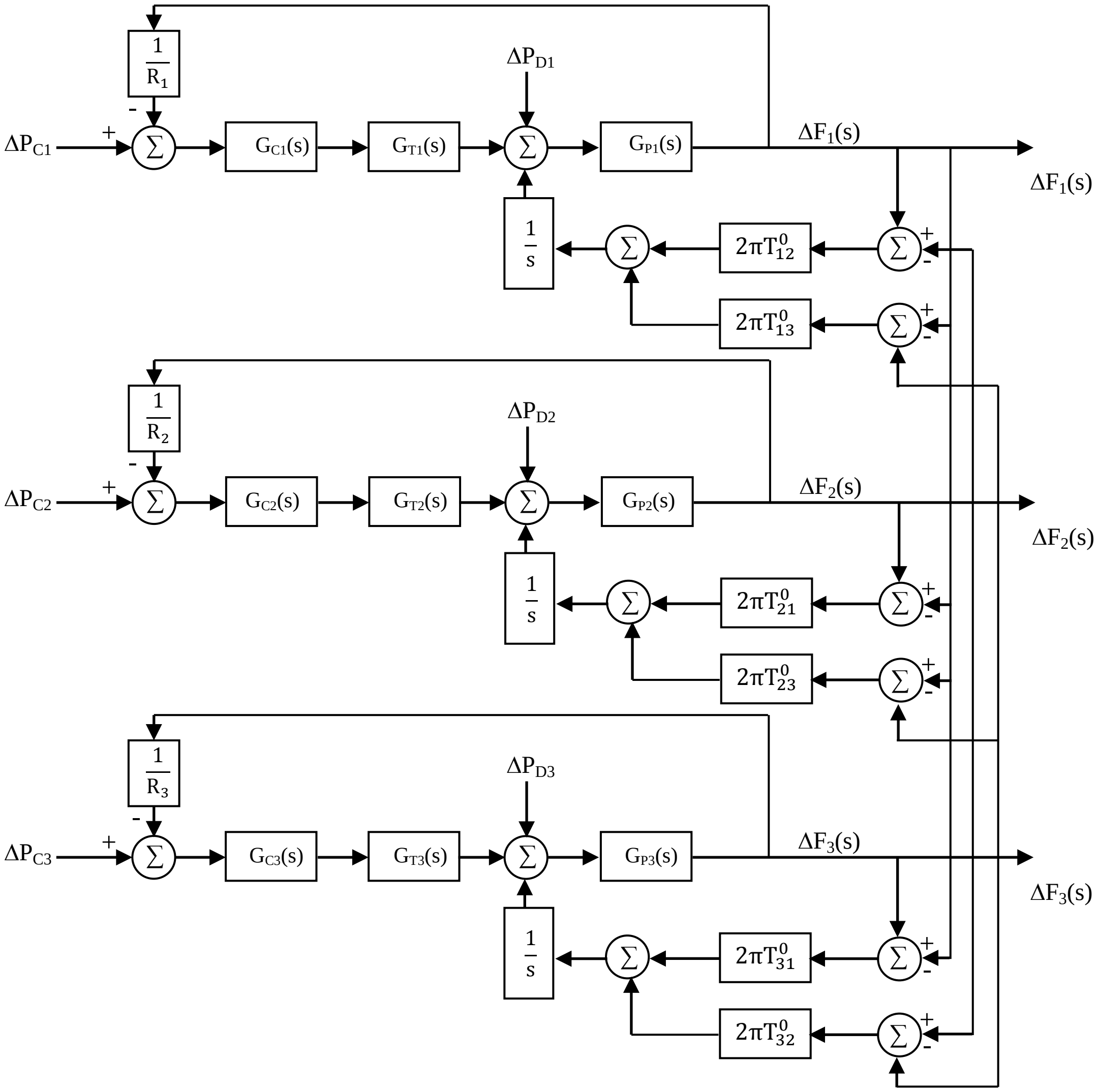

Figure 1. Three-area load-frequency control block diagram [8].

$$\Delta P_{tie,3} = \frac{2\pi T_{31}^{0}}{s}[\Delta F_3(s) - \Delta F_2(s)] + \frac{2\pi T_{32}^{0}}{s}[\Delta F_3(s) - \Delta F_2(s)] \quad (12)$$

These can be converted into time-domain state equations as derived in [8],

$$\Delta \dot{f}_1 = \frac{1}{T_{P_1}}\left(-\Delta f_1 + K_{P1}\Delta P_{G1} - K_{P1}\Delta P_{D1} - K_{P1}\Delta P_{tie,1}\right) \quad (13)$$

$$\Delta \dot{f}_2 = \frac{1}{T_{P_2}}\left(-\Delta f_2 + K_{P2}\Delta P_{G2} - K_{P2}\Delta P_{D2} - K_{P2}\Delta P_{tie,2}\right) \quad (14)$$

$$\Delta \dot{f}_3 = \frac{1}{T_{P_3}}\left(-\Delta f_3 + K_{P3}\Delta P_{G3} - K_{P3}\Delta P_{D3} - K_{P3}\Delta P_{tie,3}\right) \quad (15)$$

$$\Delta \dot{P}_{G1} = \frac{1}{T_{T_1}}\left(-\Delta P_{G1} + \Delta X_{V1}\right) \quad (16)$$

$$\Delta \dot{P}_{G2} = \frac{1}{T_{T_2}}\left(-\Delta P_{G2} + \Delta X_{V2}\right) \quad (17)$$

$$\Delta \dot{P}_{G3} = \frac{1}{T_{T_3}}\left(-\Delta P_{G3} + \Delta X_{V3}\right) \quad (18)$$

$$\Delta \dot{X}_{V1} = \frac{1}{T_{S1}}\left(-\Delta X_{V1} + \Delta P_{C1} - \frac{1}{R_1}\Delta f_1\right) \quad (19)$$

$$\Delta \dot{X}_{V2} = \frac{1}{T_{S2}}\left(-\Delta X_{V2} + \Delta P_{C2} - \frac{1}{R_2}\Delta f_2\right) \quad (20)$$

$$\Delta \dot{X}_{V3} = \frac{1}{T_{S3}}\left(-\Delta X_{V3} + \Delta P_{C3} - \frac{1}{R_3}\Delta f_3\right) \quad (21)$$

$$\Delta \dot{P}_{tie,1} = 2\pi T_{12}^{0}(\Delta f_1 - \Delta f_2) + 2\pi T_{13}^{0}(\Delta f_1 - \Delta f_3) \quad (22)$$

$$\Delta \dot{P}_{tie,2} = 2\pi T_{21}^{0}(\Delta f_2 - \Delta f_1) + 2\pi T_{23}^{0}(\Delta f_2 - \Delta f_3) \quad (23)$$

$$\Delta \dot{P}_{tie,3} = 2\pi T_{31}^{0}(\Delta f_3 - \Delta f_1) + 2\pi T_{32}^{0}(\Delta f_3 - \Delta f_2) \quad (24)$$

These can be more compactly written in matrix-vector form as,

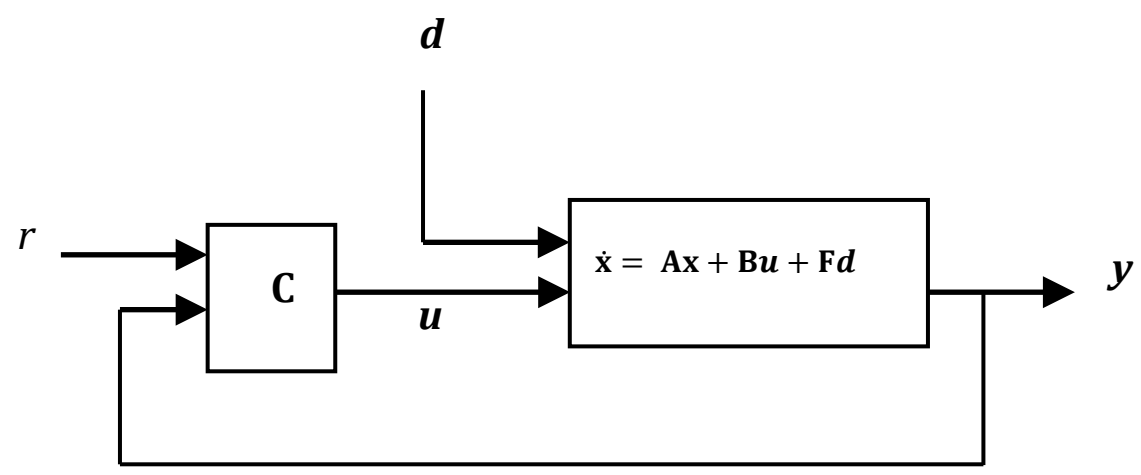

Figure 2. Optimal controller configuration.

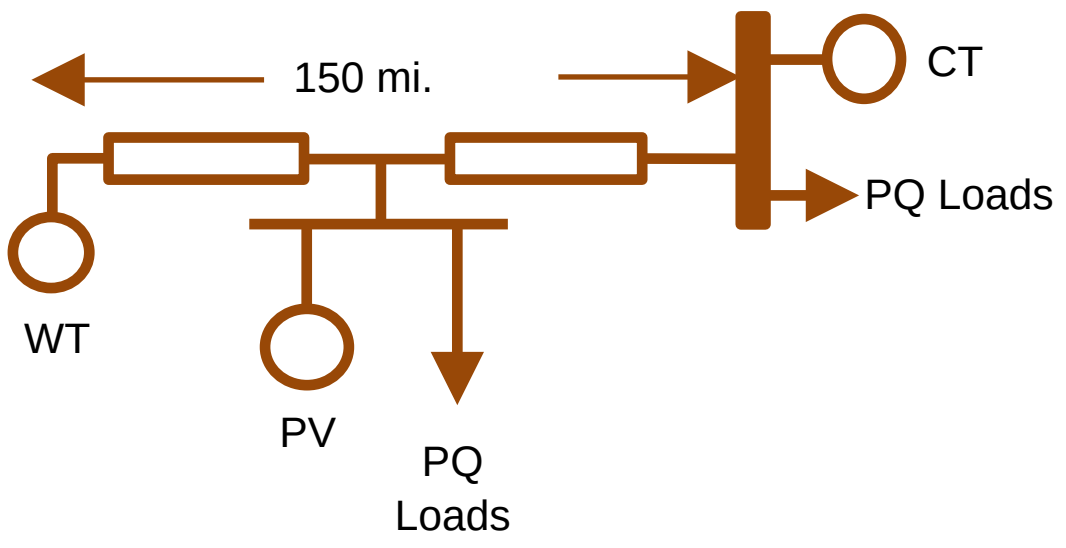

Figure 3. Three-area load-frequency control application.

$$\dot{\mathbf{x}} = \mathbf{Ax} + \mathbf{Bu} + \mathbf{Fd} \quad (25)$$

where the state vector is

$$x^T = \left[\Delta f_1\ \Delta f_2\ \Delta f_3\ \Delta P_{G1}\ \Delta P_{G2}\ \Delta P_{G3}\ \Delta P_{V1}\ \Delta P_{V2}\ \Delta P_{V3}\ \Delta P_{tie,1}\ \Delta P_{tie,2}\ \Delta P_{tie,3}\right] \quad (26)$$

And the control inputs are the commanded power provided by each source,

$$u^T = [\Delta P_{C1}\ \Delta P_{C2}\ \Delta P_{C3}]\,, \quad (27)$$

and the system responds to changes in demanded loads as disturbance inputs to the system,

$$d^T = [\Delta P_{D1}\ \Delta P_{D2}\ \Delta P_{D3}]\,, \quad (28)$$

with measurable states that can be used for feedback control purposes,

$$y^T = \left[\Delta f_1\ \Delta f_2\ \Delta f_3\ \Delta P_{tie,1}\ \Delta P_{tie,2}\ \Delta P_{tie,3}\right]. \quad (29)$$

The availability of (29) for feedback control is justified by the specifications for IEEE C37.118 compliant phasor measurement units where time latency and accuracy are minimized [12].

## III. Stochastic Optimal Control

The statistical characteristics of the disturbance inputs of wind and solar generating levels can be collected and analyzed. Given the mean and variance calculations from the generating historical data, a linear quadratic Gaussian (LQG) controller can be developed [10]. The controller includes a Kalman filter that estimates the system state variables that are not directly measureable as $\hat{x}$. In order to reduce the steady-state frequency to zero (that is, to provide frequency regulation capability for each area), the integral if each area's frequency error is defined as state augmented variables $x_i$,

$$\begin{bmatrix}\dot{\hat{x}}\\ \dot{x}_i\end{bmatrix} = \begin{bmatrix}A - BK_x - LC + LDK_x & -BK_i + LDK_i\\ 0 & 0\end{bmatrix}\begin{bmatrix}\hat{x}\\ x_i\end{bmatrix} + \begin{bmatrix}0 & L\\ I & -I\end{bmatrix}\begin{bmatrix}r\\ y\end{bmatrix} \quad (30)$$

The controller is defined as a proportional-integral PI compensator,

$$u = [-K_x \quad -K_i]\begin{bmatrix}\hat{x}\\ x_i\end{bmatrix} \quad (31)$$

The LQG controller employs statistical characteristics of the system in order to develop the optimal state estimate and controller gains [6]. The statistical model is derived using historical data collected from the operational PMU data that has been collected from the same system that requires stabilization. A block diagram showing the controller LQG controller structure is shown in Fig. 2.

## IV. Evaluation of Optimal Load-Frequency Control

Parameter values are selected that correspond to a three-area system as shown in Fig. 3. Area-1 consists of 600 MW wind-turbine generators (WTG) with Type-4 doubly-fed induction generators. Area 2 consists of local loads along with 400 MW of solar photovoltaic (PV) generation with local real and reactive (PQ) loads. Area 3 has 800 MW of conventional combustion turbines (CT) and local PQ loads. The three areas are interconnected with long transmission lines. The effects of PV fluctuations are examined with and without the LQG load-frequency control of the static inverters located in Area 1 and Area 2. Figures 4 and 5 indicate the frequency and intertie-line transients respectively, in the wind turbine area (WTG) Area-1 after a 50% reduction in rated solar PV power in Area 2. It is observed that the LQG controller provides significantly better regulation in the frequency and power stability performance in the neighboring WTG area. The corresponding transient in intertie-line power and frequency of Area-3 is shown in Figs. 6 and 7, respectively. The LQG controller provides improved damping of the inter-area power flows as well as improved stability of the CT frequency.

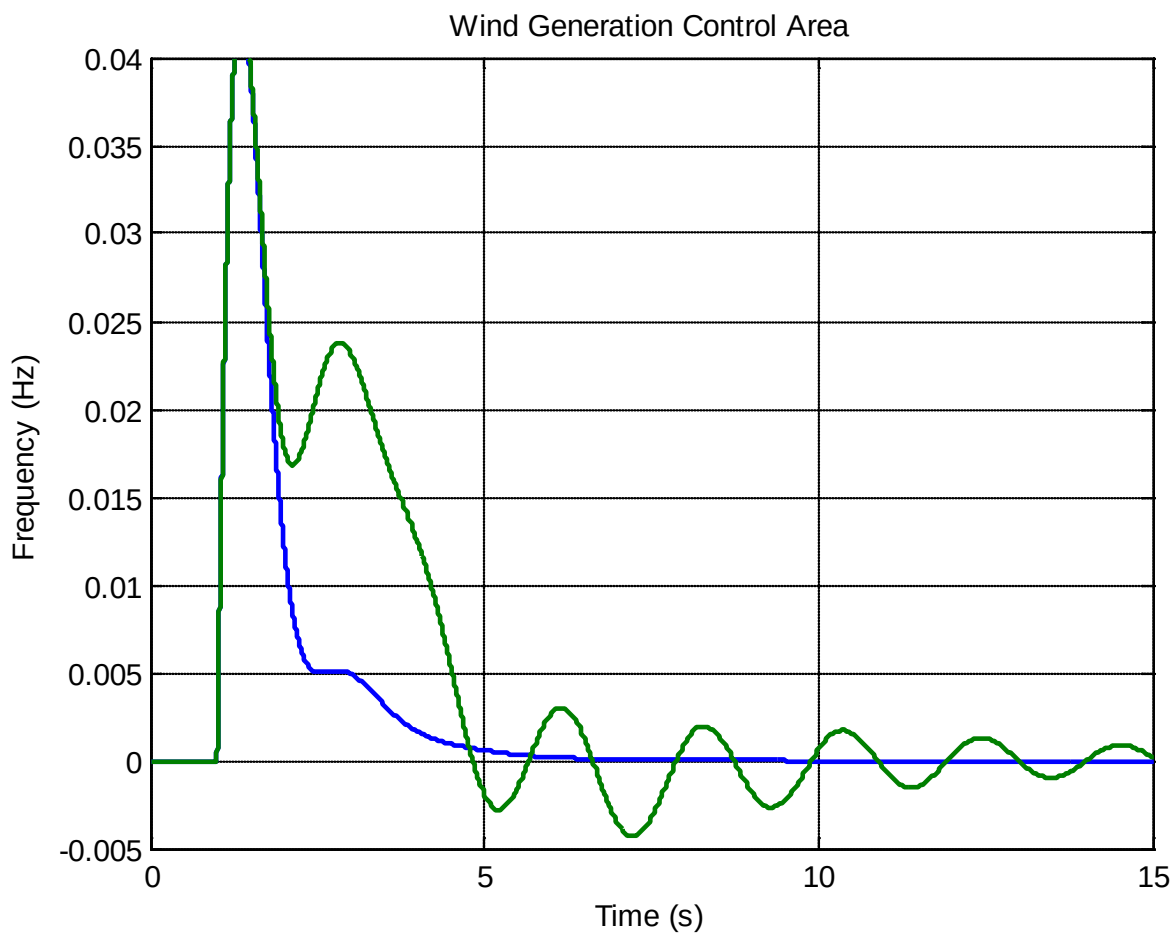

Figure 4. WTG frequency transient due to 50% PV reduction in Area-2.

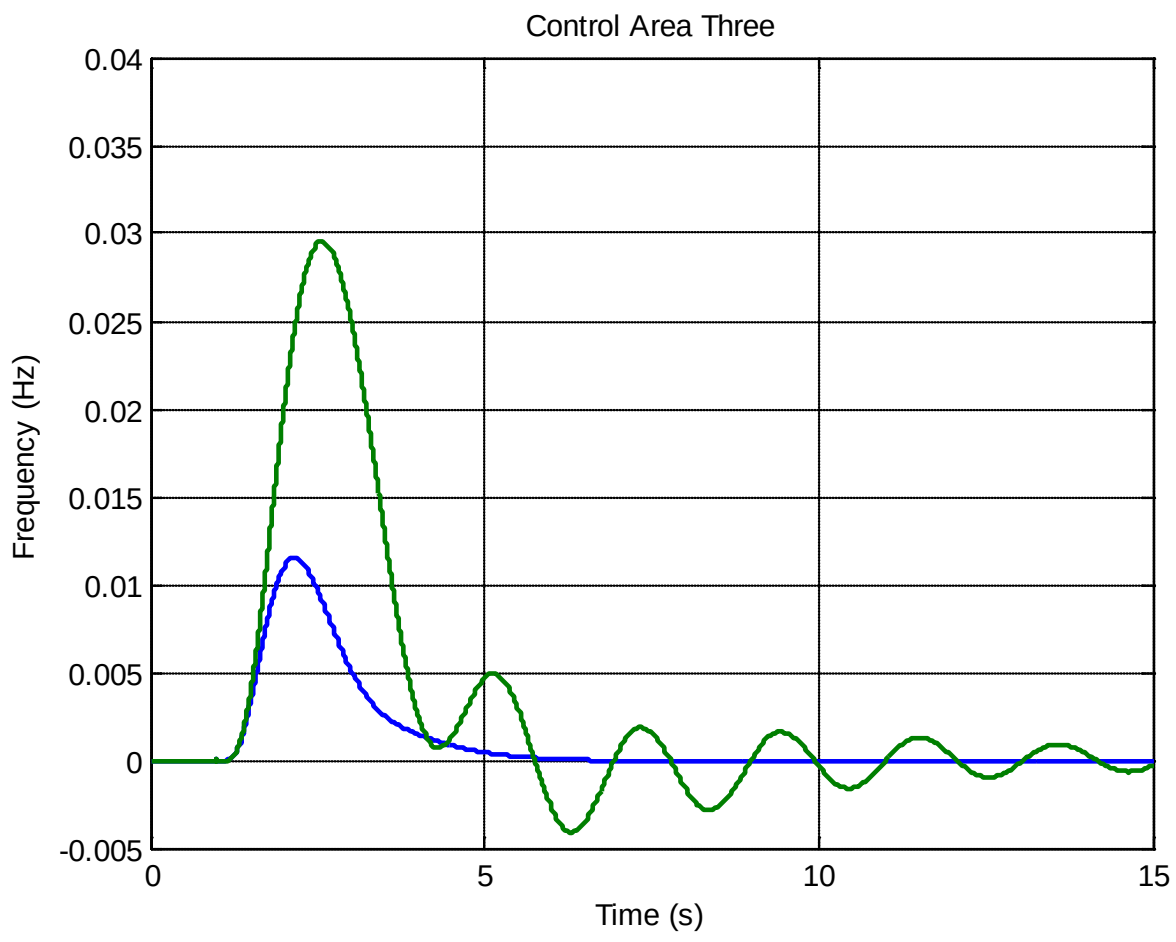

Figure 6. CT frequency transient due to 50% PV reduction in Area-2.

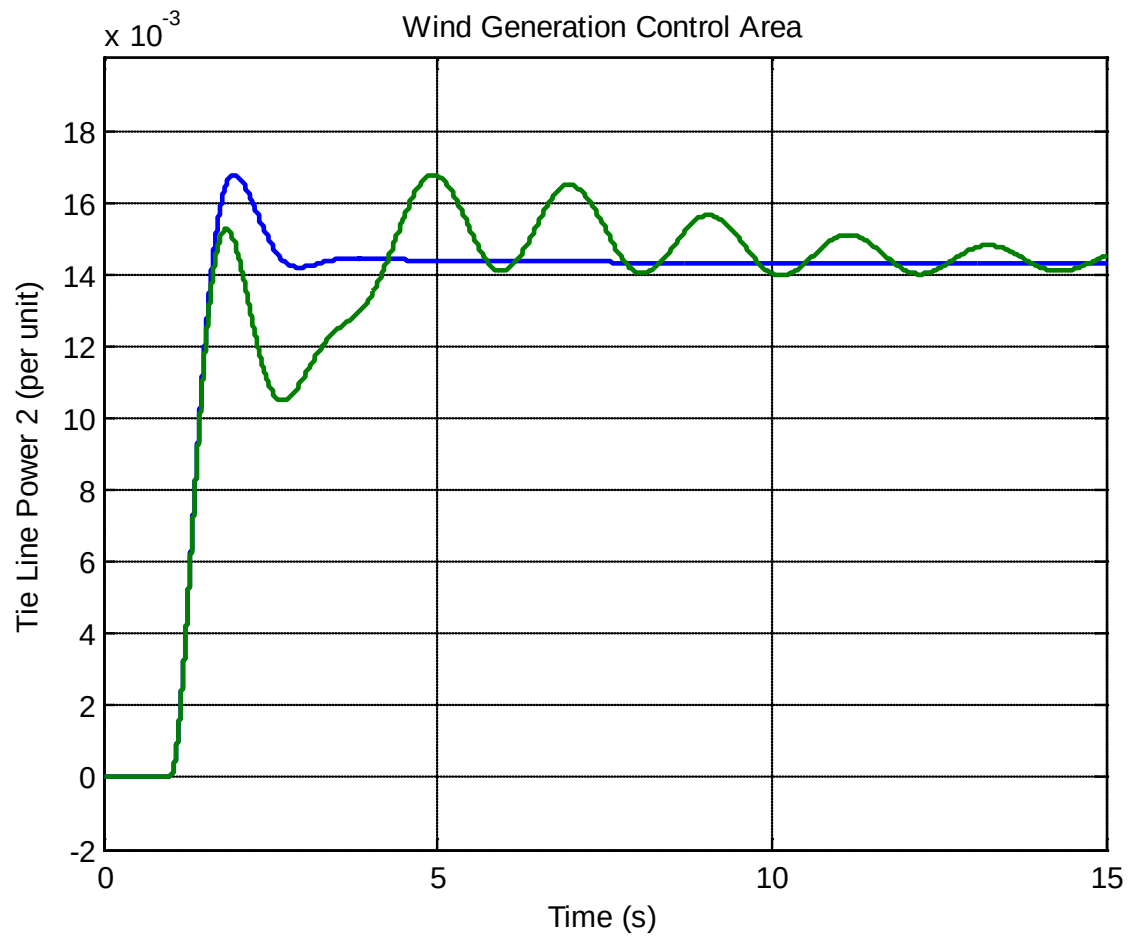

Figure 5. WTG intertie-line power due to 50% PV reduction in Area-2.

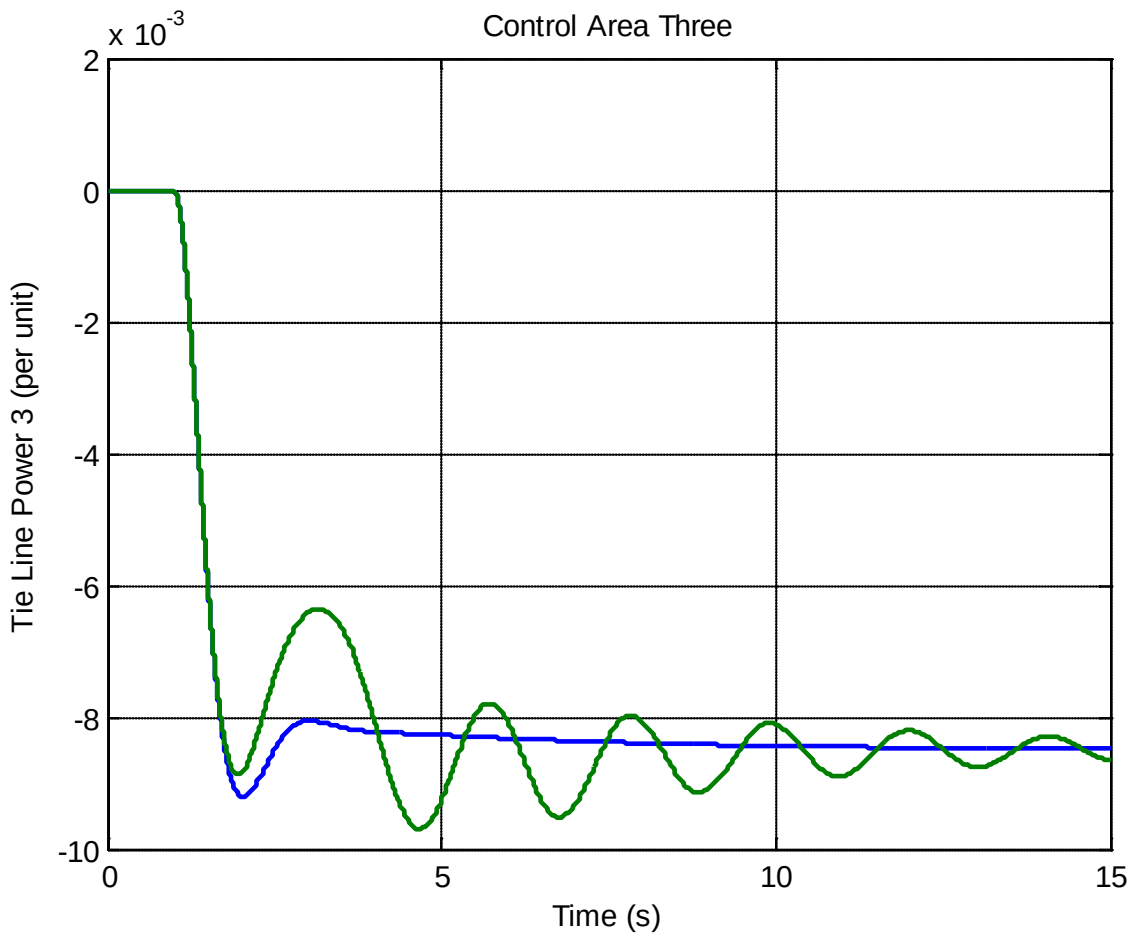

Figure 7. CT intertie-line power transient due to 50% PV reduction in Area-2.

## V. Conclusions

A method for modeling multi-area load-frequency dynamics is provided that accounts for the dynamics of static inverters. The frequency droop characteristics that are defined for conventional synchronous generators are adapted to power electronic inverters with a stochastic optimal control method that includes the faster time constants associated with most inverter circuits. Simulation results confirm the benefits of this method for a three area load-frequency control system where improved ability to include renewable energy sources is demonstrated.